%% file: main.tex
\documentclass[sigconf,nonacm]{acmart}
\usepackage{enumitem}
\AtBeginDocument{%
  }

\hypersetup{
    colorlinks=true,
    urlcolor=blue,
}

\usepackage{xcolor}

\begin{document}

\title{SPRITE: From Static Mockups to Engine-Ready Game UI}

\author{Yunshu Bai}
\affiliation{%
  \institution{Shanghai University}
  \city{Shanghai}
  \country{China}}
\email{baisherrywhite@shu.edu.cn}
\affiliation{%
  \institution{MiAO Worlds}
  \city{Singapore}
  \country{Singapore}}
\email{cielobai@miao.company}

\author{RuiHao Li}
\affiliation{%
  \institution{MiAO Worlds}
  \city{Singapore}
  \country{Singapore}}
\email{rioli@miao.company}

\author{Hao Zhang}
\affiliation{%
  \institution{MiAO Worlds}
  \city{Singapore}
  \country{Singapore}}
\email{haozhang@miao.company}

\author{Chien Her Lim}
\affiliation{%
  \institution{MiAO Worlds}
  \city{Singapore}
  \country{Singapore}}
\email{limchienher@miao.company}

\author{Ming Yan}
\authornote{Corresponding authors.}
\affiliation{%
  \institution{MiAO Worlds}
  \city{Singapore}
  \country{Singapore}}
\email{mingyan@miao.company}

\author{Mengtian Li}
\authornotemark[1]
\affiliation{%
  \institution{Shanghai University}
  \city{Shanghai}
  \country{China}}
\email{mtli@shu.edu.cn}

\renewcommand{\shortauthors}{Yunshu Bai et al.}

\input{sec/0_abstract}  

\ccsdesc[500]{Human-centered computing~Systems and tools for interaction design}
\ccsdesc[500]{Human-centered computing~User interface programming}

\keywords{Game UI, Automated Refactoring, Vision-Language Models, Generative AI}

\begin{teaserfigure}
  \includegraphics[width=\textwidth]{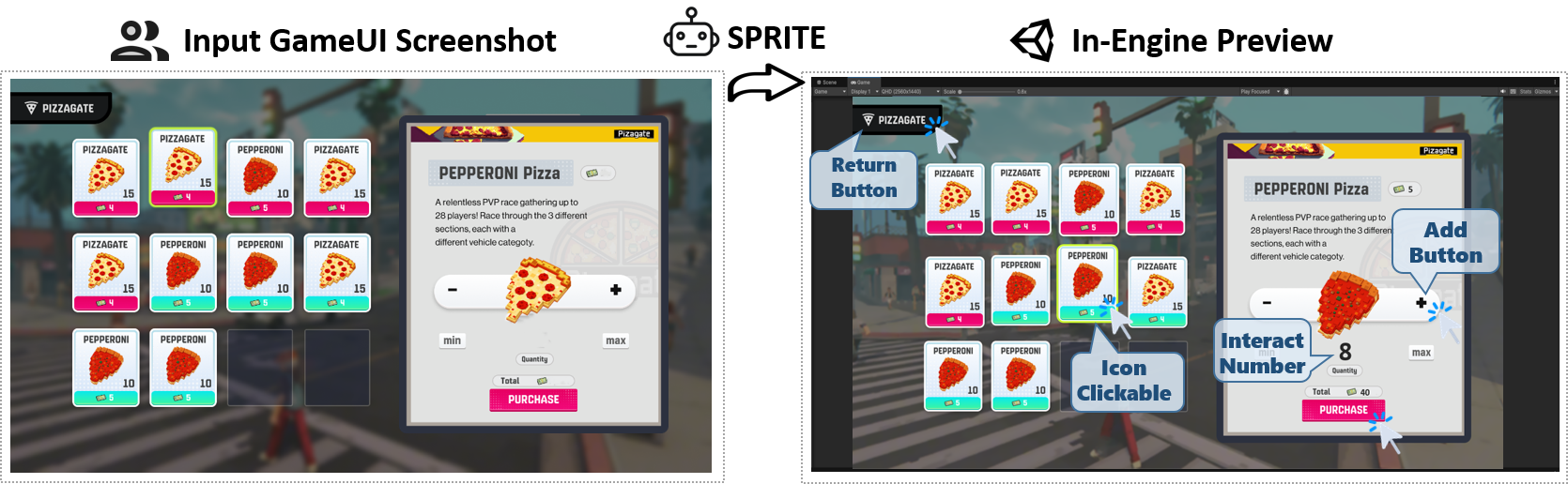}
   \Description{A pipeline overview of the SPRITE system. The left side shows a raw, highly stylized game user interface screenshot with ornate buttons and HUD elements. An arrow points to the right side, which displays the reconstructed version inside a game engine editor. The reconstruction highlights editable components, irregular geometries, and layout structures, demonstrating the automated transition from a static image to functional engine assets.}
  \caption{The SPRITE System.Transforming a raw GameUI screenshot (left) into editable engine assets (right). Unlike standard app interfaces, Game UIs are characterized by highly stylized aesthetics, irregular geometries, and complex visual hierarchies, making manual reconstruction a labor-intensive task that requires specialized expertise. By automating this reconstruction, SPRITE effectively bridges the gap between artistic vision and engine implementation to enable rapid gameplay iteration.}
  \label{fig:teaser}
\end{teaserfigure}

\maketitle
\input{sec/1_intro}

\input{sec/2_related}  
\input{sec/3_method}

\input{sec/4_Evaluation}  
\input{sec/5_conclusion}  


\bibliographystyle{ACM-Reference-Format}
\bibliography{main}

\end{document}

%% file: sec/0_abstract.tex
\begin{abstract}
Game UI implementation requires translating stylized mockups into interactive engine entities. However, current ``Screenshot-to-Code'' tools often struggle with the irregular geometries and deep visual hierarchies typical of game interfaces. To bridge this gap, we introduce SPRITE, a pipeline that transforms static screenshots into editable engine assets. By integrating Vision-Language Models (VLMs) with a structured YAML intermediate representation, SPRITE explicitly captures complex container relationships and non-rectangular layouts. We evaluated SPRITE against a curated Game UI benchmark and conducted expert reviews with professional developers to assess reconstruction fidelity and prototyping efficiency. Our findings demonstrate that SPRITE streamlines development by automating tedious coding and resolving complex nesting. By facilitating rapid in-engine iteration, SPRITE effectively blurs the boundaries between artistic design and technical implementation in game development.Project page: \href{https://baiyunshu.github.io/sprite.github.io/}{\textcolor{blue}{https://baiyunshu.github.io/sprite.github.io/}}
\end{abstract}

%% file: sec/1_intro.tex
\section{Introduction}
\label{sec:intro}
Developing a Game User Interface (UI) is fundamentally an act of translation: developers must mentally map highly stylized, artistic visions into the rigid, logical structures of a game engine. Unlike utilitarian web or mobile apps, game interfaces serve as immersive narrative devices, often employing irregular geometries, diegetic elements, and deep visual hierarchies to sustain the player's suspension of disbelief. However, the cognitive load of this translation is immense. As with other forms of craftsmanship, the gap between visualizing a design and implementing it in an engine remains a persistent bottleneck. Developers are currently forced to manually slice, measure, and reconstruct assets—a laborious "design-to-engine" grind that stifles creative iteration and often leads to visual discrepancies ~\cite{tang2024autogameui}.

While ``Screenshot-to-Code'' technologies have shown promise in automating this process for standard web and mobile domains~\cite{beltramelli2018pix2code,feng2021guis2code}, they operate on a fundamental mismatch when applied to games. Mainstream approaches, ranging from seminal Deep Learning models to recent methods based on Large Language Models (LLMs) such as GPT-4V and Gemini~\cite{si2025design2code,zhang2025widget2code}, are primarily trained on standardized datasets like RICO and PubLayNet~\cite{Deka2017RicoAM,zhong2019publaynet}. These models assume a world of rectangular DOM elements and flow-based layouts. Attempting to adapt these web-centric models with modest extensions falls short because game engines rely on absolute spatial coordinates, deep custom scene graphs, and overlapping alpha-channeled sprites that lack direct HTML/CSS equivalents. Game UIs further defy this structural logic, characterized by highly stylized, non-standard widgets~\cite{Bunian2021VINSVS} and intricate, custom hierarchies that fundamentally diverge from the responsive logic of web interfaces~\cite{tang2024autogameui}. Consequently, simply generating HTML code is insufficient for game engines, which require proprietary formats (e.g., Unity UXML) and precise spatial coordinates. Bridging this gap to bring the efficiency of automated reconstruction to the highly stylized, structurally complex domain of Game UI development thus presents a significant technical challenge.

Towards addressing this semantic and structural gap, we created \textbf{SPRITE}\footnote{SPRITE stands for \textbf{S}creenshot \textbf{P}arsing and \textbf{R}econstruction of \textbf{I}nterfaces via \textbf{T}raining-free \textbf{E}ngineering. Additionally, the name alludes to the "Sprite" asset type, a fundamental 2D graphic primitive used in the Unity engine.}, a novel pipeline that operationalizes the translation from static mockups to engine-native assets. We postulated that by combining the semantic reasoning of Vision-Language Models (VLMs) with the precise localization of 2D foundation models, we could automate this reconstruction without sacrificing artistic fidelity. SPRITE employs a coarse-to-fine perception pipeline to parse non-standard elements and utilizes a YAML-based intermediate representation as a structural scaffold. This scaffold enforces hierarchical consistency, acting as a bridge that ensures visual hierarchies are precisely translated into the proprietary formats required by game engines. To evaluate SPRITE, we curated a professional-grade benchmark and conducted expert reviews with experienced developers. Our findings indicate that SPRITE’s structure-aware workflow not only accelerates the iteration loop for professional game developers by eliminating manual implementation bottlenecks, but also lowers the technical barrier for non-specialists. By automating the translation of irregular assets and complex hierarchies, SPRITE empowers a broader spectrum of creators to engage in rapid, in-engine prototyping.

In summary, our contributions are as follows:
\begin{itemize}
    \item \textbf{Instant UI Refactoring System:} SPRITE, a training-free pipeline leveraging a coarse-to-fine perception pipeline to instantly transform static screenshots into editable, engine-native assets.
    
    \item \textbf{Intermediate Bridge (Image-to-Engine):} We introduce a structure-preserving YAML schema as a semantic bridge, enforcing hierarchical consistency during code synthesis.
    
    \item \textbf{Professional-Grade Benchmark:} A high-fidelity dataset derived from industrial production environments, enabling rigorous evaluation of layout accuracy and usability against expert standards.
\end{itemize}

%% file: sec/2_related.tex
\section{Related Work}
\label{sec:Relatedwork}
\subsection{AI-Driven UI-to-Code Generation}

While code generation from natural language has advanced rapidly~\cite{Le2020DeepLF,Li2023StarCoderMT}, generating executable code directly from visual inputs remains a challenging frontier. 
Early approaches relied on heuristic computer vision techniques~\cite{Nguyen2015ReverseEM} or deep learning encoders~\cite{beltramelli2018pix2code,Soselia2023LearningUR,Xu2021image2emmetAC}, but the field has recently shifted towards Multimodal Large Language Model (MLLM)-driven pipelines. 
To mitigate common MLLM hallucinations—such as element misclassification and layout misalignment—recent strategies employ divide-and-conquer prompting~\cite{Wan2024AutomaticallyGU,Wu2025MLLMBasedUA,Zhou2025DeclarUIBD,Chi2025PluginFS} and domain-specific fine-tuning~\cite{Jiang2025ScreenCoderAV,Yang2025UIUGAU,Laurenon2024UnlockingTC}. 
However, despite satisfactory performance in standardized web and mobile environments, these methods often yield simplistic outputs when applied to complex game scenarios~\cite{si2025design2code}. 
Our approach addresses this limitation by specifically targeting the unique challenges of game interface reconstruction: irregular widget geometries and deep, custom scene graphs that remain intractable for standard layout analysis tools.

\subsection{Fragmentation in Game UI/UX Engineering}

Existing approaches typically bifurcate into visual layout generation versus logical structure correspondence. Visual-centric research~\cite{Gupta2020LayoutTransformerLG, Inoue2023LayoutDMDD}, including diffusion models~\cite{wei2024aiinspired}, synthesizes high-fidelity aesthetics but often treats layouts as flattened images, overlooking the deep view hierarchies essential for engines. Conversely, logic-centric methods~\cite{Patil2020LayoutGMNNG,Xu2022HierarchicalLB} excel at structural matching but often sacrifice the pixel-perfect fidelity needed for production. Recent efforts attempt to bridge these streams. Web-based MLLM tools~\cite{si2025design2code,Lu2024MistyUP} lack the structural rigidity required for games, while AutoGameUI~\cite{tang2024autogameui} relies on often-unavailable paired design files. In contrast, SPRITE achieves direct reconstruction from a  screenshot, generating engine-native assets without pre-aligned data.

\begin{figure*}[t]
  \centering
  \Description{SPRITE pipeline: (1) VLM converts mockup to YAML; (2) 2D models extract and calibrate elements; (3) MLLM generates UXML/USS and interaction logic.}
  \includegraphics[width=\linewidth]{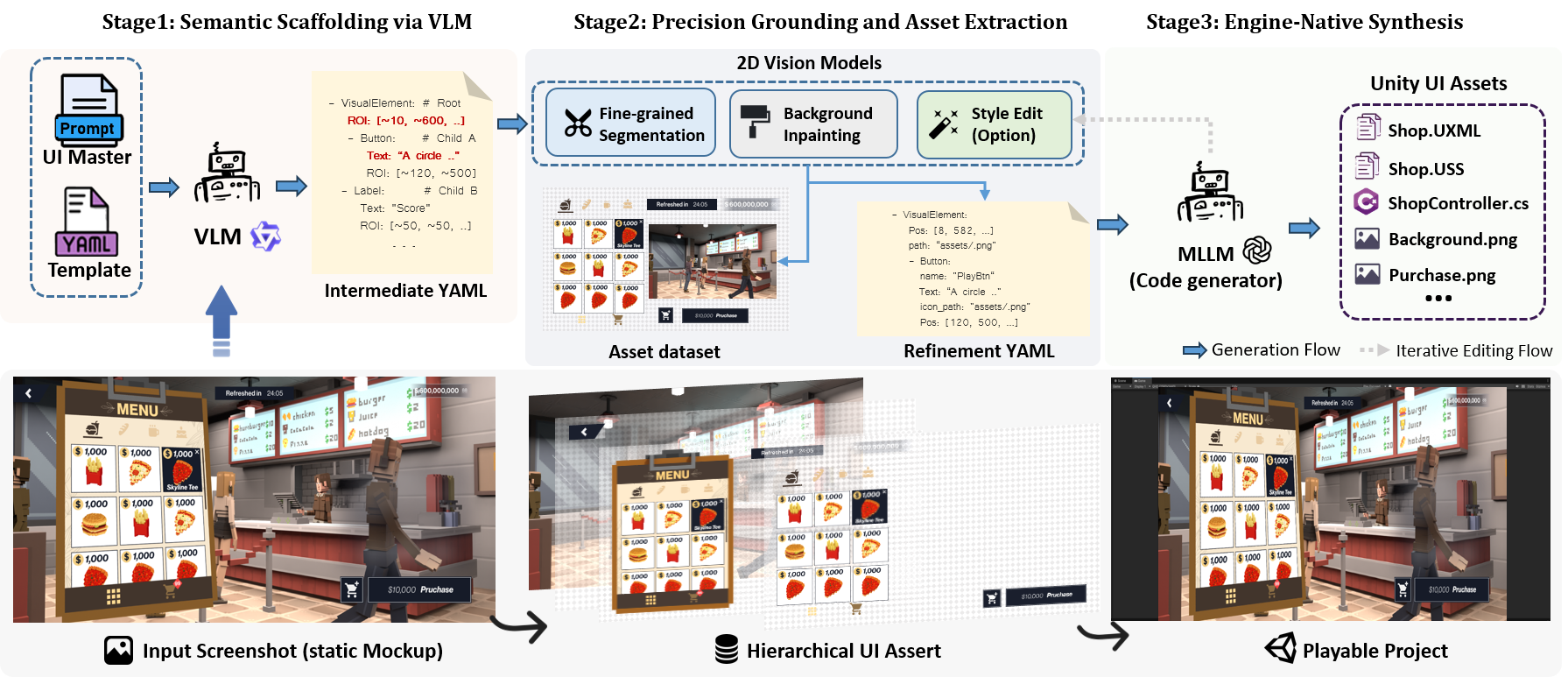}
  \caption{\textbf{SPRITE} Our system transforms mockups into engine assets via three stages: (1) \textbf{Semantic Scaffolding}, VLM infers a schema-guided hierarchical YAML; (2) \textbf{Precision Grounding}, utilizing 2D models for pixel-perfect extraction and geometric calibration; and (3) \textbf{Engine-Native Synthesis}, where an MLLM generates executable UXML/USS code and interaction logic.}
 \label{fig:pipeline}
\end{figure*}

\subsection{Data Constraints and the Shift to Training-Free Alignment}

The efficacy of UI modeling has traditionally relied on the \textit{supervised learning} paradigm, training Transformer-based models to map latent representations to layouts~\cite{Gupta2020LayoutTransformerLG,Inoue2023LayoutDMDD}. 
However, this approach is fundamentally bounded by data availability. Public datasets like RICO~\cite{Deka2017RicoAM} and PubLayNet~\cite{zhong2019publaynet} provide only flattened RGB screenshots, lacking the raw RGBA layers and complex occlusion relationships inherent in game rendering pipelines. 
To bypass this scarcity, earlier studies employed \textit{heuristic optimization}~\cite{Kumar2011BricolageER,Brisset2021ErratumLF,Kuhn1955TheHM} (e.g., integer programming~\cite{Dayama2021InteractiveLT, Xu2022HierarchicalLB}) to align visual elements with structures. While these training-free methods avoid data dependencies, their generalization is limited by rigid, handcrafted constraints. Addressing these limitations, our work adopts a \textit{training-free} VLM-driven approach. 
Instead of relying on brittle heuristics or supervised metric learning~\cite{Patil2020LayoutGMNNG, Patil2019READRA}, we leverage the inherent semantic reasoning of state-of-the-art MLLMs~\cite{anthropic2025sonnet45, Bai2025Qwen3VLTR}—a zero-shot paradigm recently proven effective in other game domains like procedural 3D map generation~\cite{her2025zeroshot3dmap} to achieve precise alignment between visual pixels and engine-native structures, effectively bypassing the bottleneck of large-scale data annotation.

%% file: sec/3_method.tex
\section{SPRITE System}
\label{sec:SPRITE System}

\subsection{Design Goals}
SPRITE aims to bridge the fundamental semantic gap between static visual intent and functional game implementation. Informed by the ``design-to-engine'' bottlenecks identified in prior formative studies~\cite{tang2024autogameui}, we formulate three design goals to guide our system:

\begin{description}[
    style=nextline, 
    leftmargin=1em,
    labelindent=0pt,
    labelsep=0.5em,
    font=\bfseries,
    itemsep=4pt
]
    \item[DG1: Fidelity Beyond Pixels: Semantic \& Structural Recovery.] 
    Game UIs are characterized by irregular geometries and diegetic elements that defy standard web-based DOM logic. Therefore, the system must go beyond mere visual replication. It must accurately interpret the \textit{semantics} of non-standard assets and reconstruct their deep, nested hierarchies, ensuring the output is a logically sound scene graph ready for engine interaction.

    \item[DG2: Zero-Shot Generalizability via Training-Free Architecture.] 
    To serve the diverse game development community, the system must be agnostic to visual styles without requiring costly domain-specific retraining. By leveraging a coarse-to-fine perception pipeline and a modular YAML intermediate representation, SPRITE aims to provide a robust, generalized bridge that adapts to various genres out-of-the-box.

    \item[DG3: Dual-Audience Empowerment: Accessibility \& Augmentation.] 
    Recognizing the varying technical expertise within the community, SPRITE is designed to support a spectrum of user needs:
    \begin{itemize}[nosep, leftmargin=1em, topsep=2pt, label={\tiny$\bullet$}] 
        \item \textit{Accessible Creation for Novices: } 
        Lowering the technical barrier to entry, enabling non-programmers to instantly transform sketches into functional prototypes.
        \item \textit{Augmenting Workflows for Professionals: } 
        Eliminating the tedious ``grunt work'' of pixel-level manual slicing and layout scaffolding, thereby freeing experts to focus on complex interaction logic and creative iteration.
    \end{itemize}

\end{description}

\subsection{The SPRITE Pipeline}
SPRITE leverages a training-free, ``coarse-to-fine'' pipeline to bridge the gap between static screenshots and functional game engines. \textbf{As illustrated in Figure~\ref{fig:pipeline}}, the pipeline decomposes complex UI reconstruction into three structured stages, minimizing manual labor while ensuring high-fidelity output (DG1, DG2).

\subsubsection{Stage 1: Semantic Scaffolding via VLM} 
To handle the non-standard and highly stylized nature of game UIs, SPRITE employs a \textbf{schema-guided reasoning} strategy. We utilize a pre-defined YAML template that mirrors the hierarchical logic of engine formats (e.g., UXML), serving as the structural contract between visual pixels and code. Driven by an \textit{Expert-Persona} prompt (``UI Master''), the VLM scans the screenshot to populate this template. It identifies functional components (e.g., HUDs, progress bars) and maps them onto a nested structure, constructing an initial \textbf{semantic scene graph} (DG2). 

\begin{figure}[htbp]
    \centering
    \Description{A code snippet illustrating a YAML structure. It shows root-level container panels with parent set to null, and nested interactive elements referencing their parent. Each element includes fields for label, 2D bounding box coordinates, and a segmentation prompt.}
    \includegraphics[width=0.8\linewidth]{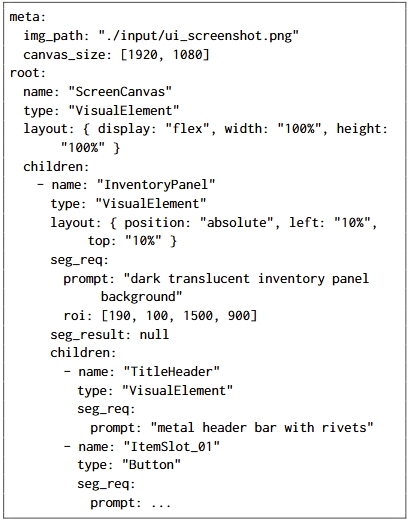} 
    \caption{YAML template structure for UI scaffolding.}
    \label{fig:yaml_template}
\end{figure}

This intermediate representation ensures that high-level logical consistency is established before the system attempts pixel-level processing. To optimize this scaffolding, we explicitly adopt YAML instead of JSON for two strategic reasons: (1) \textbf{Token Efficiency:} YAML eliminates redundant syntactic overhead (e.g., extensive braces), reducing token consumption by approximately 20--30\% and allowing the VLM to process denser layouts; (2) \textbf{UXML Semantic Alignment:} YAML's indentation-based nesting naturally mirrors the structure of Unity's UXML, ensuring structural isomorphism for downstream engine synthesis. See the template in Figure~\ref{fig:yaml_template}.

\subsubsection{Stage 2: Precision Grounding and Asset Extraction} 
While VLMs excel at semantics, they often struggle with spatial precision. To mitigate this, SPRITE integrates 2D foundation models for geometric refinement (DG1). 
Guided by the Regions of Interest (ROI) defined in the YAML, the system performs \textbf{fine-grained segmentation} to extract standalone assets with alpha channels. Crucially, to ensure ``engine-ready'' completeness, an inpainting model performs \textbf{occlusion recovery}, filling background gaps left by extracted layers. The system then backfills the YAML template with precise bounding box coordinates and asset paths, effectively transitioning the scaffold from ``fuzzy semantics'' to ``exact geometry''.

\begin{figure}[t]
    \centering
    \includegraphics[width=0.85\linewidth]{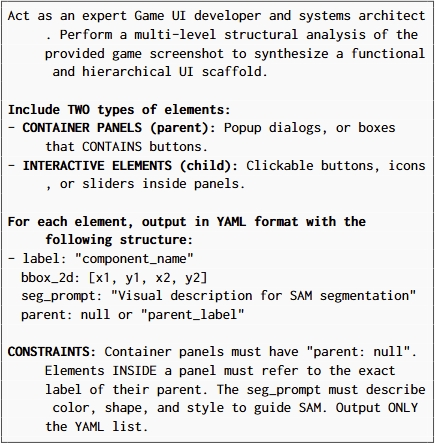} 

    \Description{A text box showing the system prompt for the UI Master Persona. It instructs the vision-language model to act as an expert Game UI developer, analyze a screenshot, and output a multi-level hierarchical UI scaffold. It specifies two types of elements: container panels and interactive elements, requiring outputs in YAML format with bounding boxes, segmentation prompts, and parent references.}
    \caption{System Prompt: UI Master Persona}
    \label{fig:ui_master_prompt}
\end{figure}

\subsubsection{Stage 3: Engine-Native Synthesis and Output} 
In the final stage, an LLM translates the calibrated YAML into executable technical assets. 
Treating the YAML as a structured specification, the model synthesizes Unity-native code (UXML/USS) that strictly adheres to the intended hierarchy. Furthermore, by interpreting semantic labels, the system infers interactive \textbf{affordances} (e.g., button hover states), enabling immediate prototyping. 
This process culminates in a \textbf{dual-value output} (DG3): providing a ``no-code'' gateway for novices to instantly realize their visions, while delivering an editable, layered ``expert scaffold'' that accelerates professional development workflows.

\begin{figure*}[t]
  \centering
  \Description{Visual comparison of four methods: (a-b) VLMs show bounding boxes; (c) baseline shows fragmented segments; (d) SPRITE extracts precise, irregular assets.}
  \includegraphics[width=\linewidth]{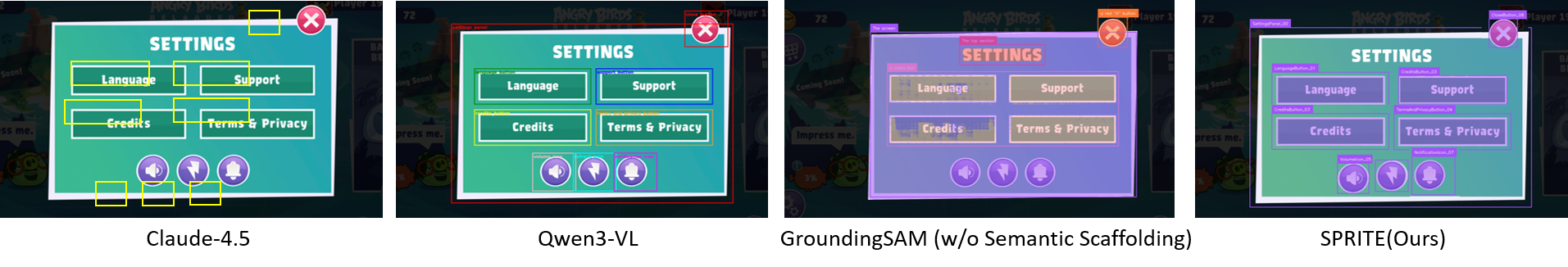}
  \caption{\textbf{Qualitative Comparison.} While VLMs (a-b) are limited to bounding boxes and the baseline (c) suffers from fragmentation, \textbf{SPRITE} (d) leverages semantic scaffolding to extract pixel-perfect assets for irregular shapes.}
   \label{fig:visual}
\end{figure*}

\subsection{Implementation Details}
SPRITE is architected as a modular, training-free pipeline that orchestrates a suite of state-of-the-art foundation models.

\subsubsection{Prompt Engineering and Visual Perception.}
For high-level semantic parsing and coarse component identification, we employ \textbf{Qwen3-VL}~\cite{Bai2025Qwen3VLTR}. This initial parsing is driven by a carefully crafted system prompt (the ``UI Master Persona''). Our prompt design follows three core rationales: (1) \textbf{Functional Decoupling} to force the VLM to filter aesthetic noise and isolate core UI logic; (2) \textbf{Structural Formalization} to mandate a \texttt{parent} field, transforming flat detection into a hierarchical scene graph; and (3) \textbf{Downstream Synergy} to generate descriptive visual cues that guide subsequent 2D foundation models. The core instruction is illustrated in Figure~\ref{fig:ui_master_prompt}.

\subsubsection{Geometric Extraction and Engine Synthesis.}
The semantic layer is augmented by a geometric stack: \textbf{GroundingDINO}~\cite{liu2023grounding} performs open-vocabulary detection based on the VLM's visual cues, while the \textbf{Segment Anything Model (SAM2)}~\cite{Ravi2024SAM2S} executes precise, pixel-level mask generation. To ensure assets are production-ready, \textbf{LaMa}~\cite{suvorov2022resolution} handles background inpainting, resolving occlusion artifacts to produce clean sprites. Finally, the core synthesis engine uses \textbf{GPT-5}~\cite{Singh2025OpenAIGS} and \textbf{Claude 4.5 Sonnet}~\cite{anthropic2025sonnet45} to translate the structured YAML into Unity UXML and USS files. We apply few-shot prompting and chain-of-thought reasoning to maintain strict hierarchical consistency. Additionally, we integrate \textbf{FLUX}~\cite{Labs2025FLUX1KF} to empower users with generative editing capabilities for stylized asset refinement.

%% file: sec/4_Evaluation.tex
\section{Evaluation}
\label{sec:Evaluation}
\subsection{GAMEUI Benchmark}

\begin{figure*}[t] 
   \centering
   \Description{Visual representation of the GAMEUI Benchmark gallery. }
   \includegraphics[width=\textwidth]{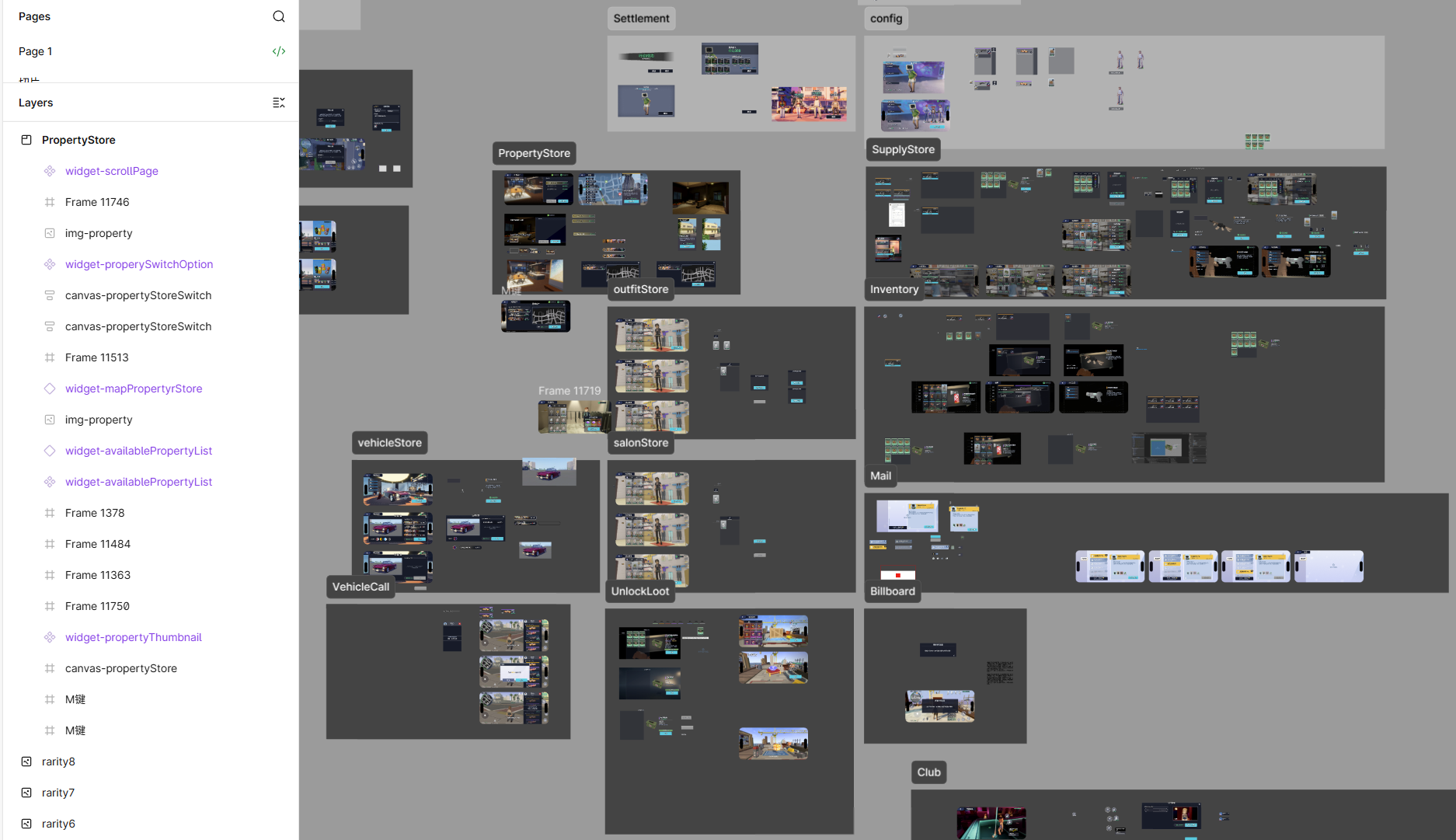} 
   \caption{Visual representation of the GAMEUI Benchmark gallery. These representative samples demonstrate the system's ability to faithfully reconstruct a wide array of challenging layouts across different game genres. Ranging from high-density RPG inventory grids to complex, multi-level navigation systems, this gallery highlights the functional diversity and zero-shot generalizability explicitly targeted by SPRITE.}
   \label{fig:gameui_dataset}
\end{figure*}

To evaluate SPRITE in realistic scenarios, we curated the GAMEUI Benchmark, a specialized dataset explicitly targeting the structural complexity of gaming (Figure~\ref{fig:gameui_dataset}). Specifically, the benchmark comprises hundreds of meticulously selected interfaces sampled from multiple distinct game genres (e.g., RPG, FPS, Strategy, Casual). These interfaces were chosen to ensure high diversity in layout complexity (ranging from simple main menus to dense inventory grids) and artistic styles. This rigorous selection process ensures that unlike general-purpose datasets, this collection features high-fidelity, production-level interfaces spanning diverse functional modules (e.g., HUDs, Settings, Inventories). 

Each entry serves as a comprehensive \textbf{ground truth ecosystem} rather than a simple screenshot, comprising:

\begin{itemize}
    \item \textbf{Figma-exported JSONs:} Detailing expert-defined structural hierarchies and interaction metadata.
    \item \textbf{Segmented Sprites:} High-quality extracted image components alongside corresponding text metadata.
    \item \textbf{Hand-authored Unity UXML/ \allowbreak USS:} Professional-grade templates serving as the synthesis gold standard.
\end{itemize}

As an ongoing project, the benchmark is \textbf{continuously expanding} to encompass more diverse UI functionalities. By processing these complex, non-standard layouts, we verified SPRITE’s ability to maintain \textbf{structural integrity (DG2)} and generate engine-ready scaffolds across a wide range of functional game UI scenarios.

\subsection{System Validation and Expert Evaluation}

\subsubsection{Qualitative Analysis of Asset Extraction}

We performed a qualitative comparison against state-of-the-art VLM detectors and zero-shot segmentation baselines. As demonstrated in Figure \ref{fig:visual}, existing methods struggle with the unique visual language of games in two distinct ways. First, standard VLM approaches (Figure \ref{fig:visual}a-b) are inherently limited to coarse bounding boxes, making them incapable of isolating assets with complex, non-rectangular boundaries. Second, while pixel-level baselines (Figure \ref{fig:visual}c) attempt to extract exact shapes, they frequently suffer from severe fragmentation. Without strong semantic constraints to resolve referencing ambiguities, these models over-rely on the latent biases of their pixel-level training distributions, fracturing cohesive UI elements into illogical, unusable debris. In contrast, SPRITE (Figure \ref{fig:visual}d) overcomes both limitations by grounding the extraction process in \textbf{semantic scaffolding}. This top-down structural guidance allows our system to reliably extract pixel-perfect assets even for highly irregular and stylized shapes, ensuring the outputs are visually clean, logically grouped, and immediately engine-ready (DG1).

\subsubsection{Expert Review and User Study}

To assess SPRITE's practical utility, we conducted an expert evaluation with three senior UI/UX designers using the GAMEUI Benchmark. During these sessions, the experts interactively tested the functional prototypes and conducted a rigorous audit of the auto-segmented UI assets, structural hierarchies, and underlying UXML source files. Evaluated on a 10-point Likert scale (1 = entirely unusable, 10 = production-ready), the system achieved strong ratings in \textbf{Visual Fidelity (8.5/10)} and \textbf{Hierarchical Logic (8.0/10)}, with the designers confirming that the reconstructed assets closely mirrored industry standards. However, \textbf{Interaction Accuracy (7.0/10)} received a comparatively lower score, prompting a deeper analysis of the system's interactive scope and boundary conditions.

\subsubsection{Interaction Scope and Failure Cases}

Currently, SPRITE successfully infers basic interactive affordances such as button hover states, toggles, and simple click events derived from semantic labels. However, the system encounters typical failure modes when processing heavily overlapping translucent elements or highly abstract, diegetic UIs where visual boundaries are ambiguous. Additionally, complex temporal logic (e.g., drag-and-drop mechanics or the non-linear motion synthesis required for production polish) remains beyond the current generation scope, directly explaining the lower interaction score observed in our expert review. Despite these boundaries, experts acknowledged the efficiency of the automated workflow, validating SPRITE as a robust prototyping tool and explicitly motivating our future focus on dynamic simulation.

%% file: sec/5_conclusion.tex
\section{Discussion and Future Work}
\label{sec:Discussion and Future Wor}
While SPRITE successfully operationalizes static reconstruction, we envision three critical trajectories to further bridge the design-to-engine gap:

\textbf{From Static Visuals to Dynamic Interactivity.} Addressing the limitation of static screenshots, future iterations of SPRITE will transition towards multi-frame temporal reasoning by leveraging advanced Video-LLMs. By analyzing short gameplay clips or UI interaction recordings, the system could autonomously extract complex temporal logic, such as non-linear animation curves, state-machine transitions, and conditional visual feedback loops (e.g., hover-to-click effects). This advancement will fundamentally shift the automated reconstruction paradigm from merely capturing the structural ``look'' of the interface to computationally replicating its dynamic ``feel'', ultimately generating ready-to-use engine animation controllers and event scripts directly from video references.

\textbf{Disentanglement for Generative Design.} We aim to evolve SPRITE into a parameterized generative design system by computationally disentangling the core structural hierarchy (UXML) from its visual presentation (USS). Mapping these decoupled components into a latent semantic space will empower developers to perform prompt-driven, global restyling—such as instantly swapping color palettes and asset motifs—without altering the underlying functional logic. This capability will facilitate rapid A/B testing, enable procedural generation of UI variants for live-ops events, and drastically reduce the friction of continuous aesthetic iteration.

\textbf{Human-AI Collaboration and Workflow Cognition.} To address the socio-technical impact of automated UI reconstruction, our future work will investigate how tools like SPRITE reshape the division of labor between artists and programmers. We plan to conduct longitudinal studies to explore how users calibrate trust in automated outputs, what new breakdowns emerge during structural refactoring, and how mixed-initiative co-creation ultimately transforms the cognitive load of game development workflows.

\section{Conclusion}
\label{sec:Conclusion}

In this work, we presented SPRITE, a training-free pipeline that helps bridge the Gulf of Execution between static artistic vision and engine-native implementation. By harmonizing VLM-based semantic scaffolding with the geometric precision of 2D foundation models and LLM code generation, SPRITE effectively automates the translation of flat mockups into functional Unity assets. Our evaluation via the GAMEUI Benchmark confirms that SPRITE not only maintains high structural integrity but also significantly streamlines the development workflow. By substantially reducing the tedious overhead of manual asset slicing and assembly, SPRITE delivers a dual impact: it lowers the technical barrier for novices to engage in rapid game creation, while enabling professional developers to better focus on complex interaction logic. Ultimately, SPRITE helps redefine the human-AI relationship in game engineering, shifting the prevailing paradigm from manual pixel reconstruction toward a collaborative framework of high-level, semantic co-creation.

%% file: main.bib
@article{tang2024autogameui,
  title={AutoGameUI: Constructing High-Fidelity Game UIs via Multimodal Learning and Interactive Web-Based Tool},
  author={Tang, Zhongliang and Tan, Mengchen and Xia, Fei and Cheng, Qingrong and Jiang, Hao and Zhang, Yongxiang},
  journal={arXiv preprint arXiv:2411.03709},
  volume={abs/2411.03709},
  year={2024},
  pages={1--9},      
  numpages={9}
}

@inproceedings{beltramelli2018pix2code,
  title={pix2code: Generating code from a graphical user interface screenshot},
  author={Beltramelli, Tony},
  booktitle={Proceedings of the ACM SIGCHI symposium on engineering interactive computing systems},
  pages={1--6},
  year={2018},
  publisher = {Association for Computing Machinery},
  address = {New York, NY, USA},
}

@inproceedings{feng2021guis2code,
    title = {{GUIS2Code}: A Computer Vision Tool to Generate Code Automatically from Graphical User Interface Sketches},
    author = {Feng, Zhen and Fang, Jiaqi and Cai, Bo and Zhang, Yingtao},
    year = {2021},
    publisher = {Springer-Verlag},
    address = {Berlin, Heidelberg},
    url = {https://doi.org/10.1007/978-3-030-86365-4_5},
    doi = {10.1007/978-3-030-86365-4_5},
    booktitle = {Proceedings of the 30th International Conference on Artificial Neural Networks ({ICANN})},
    pages = {53--65},
    numpages = {13},
    location = {Bratislava, Slovakia}
}

@inproceedings{si2025design2code,
  title={Design2code: Benchmarking multimodal code generation for automated front-end engineering},
  author={Si, Chenglei and Zhang, Yanzhe and Li, Ryan and Yang, Zhengyuan and Liu, Ruibo and Yang, Diyi},
  booktitle={Proceedings of the 2025 Conference of the Nations of the Americas Chapter of the Association for Computational Linguistics: Human Language Technologies (Volume 1: Long Papers)},
  pages={3956--3974},
  year={2025},
  publisher={Association for Computational Linguistics},
  address={Albuquerque, New Mexico, USA}
}

@article{zhang2025widget2code,
  title={Widget2Code: From Visual Widgets to UI Code via Multimodal LLMs},
  author={Zhang, Houston H and Zhang, Tao and Lin, Baoze and Xue, Yuanqi and Zhu, Yincheng and Liu, Huan and Gu, Li and Ye, Linfeng and Wang, Ziqiang and Zuo, Xinxin and others},
  journal={arXiv preprint arXiv:2512.19918},
  volume={2512.19918},
  year={2025},
  pages = {1-25},
  numpages = {25},
}

@article{Bai2025Qwen3VLTR,
  title={Qwen3-VL Technical Report},
  author={Shuai Bai and Yuxuan Cai and Ruizhe Chen and Keqin Chen and Xiong-Hui Chen and Zesen Cheng and Lianghao Deng and Wei Ding and Rongyao Fang and Chang Gao and Chunjiang Ge and Wenbin Ge and Zhifang Guo and Qidong Huang and Jie Huang and Fei Huang and Binyuan Hui and Shutong Jiang and Zhaohai Li and Mingsheng Li and Mei Li and Kaixin Li and Zicheng Lin and Junyang Lin and Xuejing Liu and Jiawei Liu and Chenglong Liu and Yang Liu and Dayiheng Liu and Shixuan Liu and Dunjie Lu and Ruilin Luo and Chenxu Lv and Rui Men and Li Ying Meng and Xuancheng Ren and Xin-yi Ren and Sibo Song and Yu-Chen Sun and Jun Tang and Jianhong Tu and Jianqiang Wan and Peng Wang and Pengfei Wang and Qiuyue Wang and Yuxuan Wang and Tianbao Xie and Yihe Xu and Haiyang Xu and Jin Xu and Zhibo Yang and Mingkun Yang and Jianxin Yang and An Yang and Bowen Yu and Fei Zhang and Hang Zhang and Xi Zhang and Botao Zheng and Humen Zhong and Jingren Zhou and Fanxi Zhou and Jingren Zhou and Yuanzhi Zhu and Keming Zhu},
  journal={ArXiv},
  year={2025},
  volume={abs/2511.21631},
  pages={1--42},
  numpages={42},
}

@article{liu2023grounding,
  title={Grounding dino: Marrying dino with grounded pre-training for open-set object detection},
  author={Liu, Shilong and Zeng, Zhaoyang and Ren, Tianhe and Li, Feng and Zhang, Hao and Yang, Jie and Li, Chunyuan and Yang, Jianwei and Su, Hang and Zhu, Jun and others},
  journal={arXiv preprint arXiv:2303.05499},
  year={2023},
  volume={abs/2303.0549},
 pages={1--33},
  numpages={33},
}

@article{Ravi2024SAM2S,
  title={SAM 2: Segment Anything in Images and Videos},
  author={Nikhila Ravi and Valentin Gabeur and Yuan-Ting Hu and Ronghang Hu and Chaitanya K. Ryali and Tengyu Ma and Haitham Khedr and Roman R{\"a}dle and Chlo{\'e} Rolland and Laura Gustafson and Eric Mintun and Junting Pan and Kalyan Vasudev Alwala and Nicolas Carion and Chao-Yuan Wu and Ross B. Girshick and Piotr Doll'ar and Christoph Feichtenhofer},
  journal={ArXiv},
  year={2024},
  volume={abs/2408.00714},
  pages={1--42},
  numpages={42},
}

@inproceedings{suvorov2022resolution,
  title={Resolution-robust large mask inpainting with fourier convolutions},
  author={Suvorov, Roman and Logacheva, Elizaveta and Mashikhin, Anton and Remizova, Anastasia and Ashukha, Arsenii and Silvestrov, Aleksei and Kong, Naejin and Goka, Harshith and Park, Kiwoong and Lempitsky, Victor},
  booktitle={Proceedings of the IEEE/CVF winter conference on applications of computer vision},
  pages={2149--2159},
  year={2022},
  publisher={IEEE/CVF},
  address={Waikoloa, HI, USA}
}

@article{Labs2025FLUX1KF,
  title={FLUX.1 Kontext: Flow Matching for In-Context Image Generation and Editing in Latent Space},
  author={Black Forest Labs and Stephen Batifol and A. Blattmann and Frederic Boesel and Saksham Consul and Cyril Diagne and Tim Dockhorn and Jack English and Zion English and Patrick Esser and Sumith Kulal and Kyle Lacey and Yam Levi and Cheng Li and Dominik Lorenz and Jonas Muller and Dustin Podell and Robin Rombach and Harry Saini and Axel Sauer and Luke Smith},
  journal={ArXiv},
  year={2025},
  volume={abs/2506.15742},
  pages={1--19},
  numpages={19},
}

@article{Singh2025OpenAIGS,
  title={Openai gpt-5 system card},
  author={Singh, Aaditya and Fry, Adam and Perelman, Adam and Tart, Adam and Ganesh, Adi and El-Kishky, Ahmed and McLaughlin, Aidan and Low, Aiden and Ostrow, AJ and Ananthram, Akhila and others},
  journal={arXiv preprint arXiv:2601.03267},
  year={2025},
  volume={abs/2601.03267},
  pages={1--61},
  numpages={61}
}

@misc{anthropic2025sonnet45,
  author = {{Anthropic}},
  title = {Introducing {Claude} {Sonnet} 4.5},
  howpublished = {\url{https://www.anthropic.com/news/claude-sonnet-4-5}},
  year = {2025},
  note = {Accessed: 2026-01-20}
}

@inproceedings{Gupta2020LayoutTransformerLG,
  title={Layouttransformer: Layout generation and completion with self-attention},
  author={Gupta, Kamal and Lazarow, Justin and Achille, Alessandro and Davis, Larry S and Mahadevan, Vijay and Shrivastava, Abhinav},
  booktitle={Proceedings of the IEEE/CVF International Conference on Computer Vision (ICCV)},
  year={2021},
  pages={1004--1014},
  numpages={11},
  publisher={IEEE/CVF},
  address={Montreal, QC, Canada}
}

@inproceedings{Inoue2023LayoutDMDD,
  title={{LayoutDM}: Discrete Diffusion Model for Controllable Layout Generation},
  author={Inoue, Naoto and Kikuchi, Kotaro and Simo-Serra, Edgar and Otani, Mayu and Yamaguchi, Kota},
  booktitle={Proceedings of the {IEEE/CVF} Conference on Computer Vision and Pattern Recognition ({CVPR})},
  year={2023},
  pages={10167--10176},
  numpages={10},
  publisher={IEEE/CVF},
  address={Vancouver, BC, Canada}
}

@inproceedings{Deka2017RicoAM,
author = {Deka, Biplab and Huang, Zifeng and Franzen, Chad and Hibschman, Joshua and Afergan, Daniel and Li, Yang and Nichols, Jeffrey and Kumar, Ranjitha},
title = {Rico: A Mobile App Dataset for Building Data-Driven Design Applications},
year = {2017},
isbn = {9781450349819},
publisher = {Association for Computing Machinery},
address = {New York, NY, USA},
url = {https://doi.org/10.1145/3126594.3126651},
doi = {10.1145/3126594.3126651},
booktitle = {Proceedings of the 30th Annual ACM Symposium on User Interface Software and Technology},
pages = {845–854},
numpages = {10},
location = {Qu\'{e}bec City, QC, Canada},
series = {UIST '17}
}

@inproceedings{zhong2019publaynet,
  title={Publaynet: largest dataset ever for document layout analysis},
  author={Zhong, Xu and Tang, Jianbin and Yepes, Antonio Jimeno},
  booktitle={2019 International conference on document analysis and recognition (ICDAR)},
  pages={1015--1022},
  year={2019},
  publisher={IEEE},
  address={Sydney, NSW, Australia}
}

@inproceedings{Kumar2011BricolageER,
author = {Kumar, Ranjitha and Talton, Jerry O. and Ahmad, Salman and Klemmer, Scott R.},
title = {Bricolage: example-based retargeting for web design},
year = {2011},
isbn = {9781450302289},
publisher = {Association for Computing Machinery},
address = {New York, NY, USA},
url = {https://doi.org/10.1145/1978942.1979262},
doi = {10.1145/1978942.1979262},
booktitle = {Proceedings of the SIGCHI Conference on Human Factors in Computing Systems},
pages = {2197–2206},
numpages = {10},
location = {Vancouver, BC, Canada},
series = {CHI '11}
}

@article{Brisset2021ErratumLF,
  title={Erratum: Leveraging Flexible Tree Matching to Repair Broken Locators in Web Automation Scripts},
  author={Sacha Brisset and Romain Rouvoy and Lionel Seinturier and Renaud Pawlak},
  journal={ArXiv},
  year={2021},
  volume={abs/2106.04916},
  pages={1--34},
  numpages={34},
}

@inproceedings{Dayama2021InteractiveLT,
author = {Dayama, Niraj Ramesh and Santala, Simo and Br\"{u}ckner, Lukas and Todi, Kashyap and Du, Jingzhou and Oulasvirta, Antti},
title = {Interactive Layout Transfer},
year = {2021},
isbn = {9781450380171},
publisher = {Association for Computing Machinery},
address = {New York, NY, USA},
url = {https://doi.org/10.1145/3397481.3450652},
doi = {10.1145/3397481.3450652},
booktitle = {Proceedings of the 26th International Conference on Intelligent User Interfaces},
pages = {70–80},
numpages = {11},
location = {College Station, TX, USA},
series = {IUI '21}
}

@article{Kuhn1955TheHM,
  title={The Hungarian method for the assignment problem},
  author={Harold W. Kuhn},
  journal={Naval Research Logistics (NRL)},
  year={1955},
  volume={2},
  number={1-2},
  pages={83--97},
  numpages={15}
}

@article{Xu2022HierarchicalLB,
  title={Hierarchical Layout Blending with Recursive Optimal Correspondence},
  author={Pengfei Xu and Yifan Li and Zhijin Yang and Weiran Shi and Hongbo Fu and Hui Huang},
  journal={ACM Transactions on Graphics (TOG)},
  year={2022},
  volume={41},
  pages={1 - 15},
}

@inproceedings{Patil2019READRA,
  title={{READ}: Recursive Autoencoders for Document Layout Generation},
  author={Akshay Gadi Patil and Omri Ben-Eliezer and Or Perel and Hadar Averbuch-Elor},
  booktitle={Proceedings of the {IEEE/CVF} Conference on Computer Vision and Pattern Recognition Workshops ({CVPRW})},
  year={2020},
  pages={2316--2325},
  numpages={10},
  publisher={IEEE},
  address={Seattle, WA, USA}
}

@inproceedings{Patil2020LayoutGMNNG,
  title={{LayoutGMN}: Neural Graph Matching for Structural Layout Similarity},
  author={Akshay Gadi Patil and Manyi Li and Matthew Fisher and Manolis Savva and Hao Zhang},
  booktitle={Proceedings of the {IEEE/CVF} Conference on Computer Vision and Pattern Recognition ({CVPR})},
  year={2021},
  pages={11043--11052},
  numpages={10},
  publisher={IEEE/CVF},
  address={Nashville, TN, USA}
}

@article{Chi2025PluginFS,
  title={Plug-in Feedback Self-adaptive Attention in CLIP for Training-free Open-Vocabulary Segmentation},
  author={Zhixiang Chi and Yanan Wu and Li Gu and Huan Liu and Ziqiang Wang and Yang Zhang and Yang Wang and Konstantinos N. Plataniotis},
  journal={ArXiv},
  year={2025},
  volume={abs/2508.20265},
  pages={1--42},
  numpages={42},
  
}

@article{Zhou2025DeclarUIBD,
  title={DeclarUI: Bridging Design and Development with Automated Declarative UI Code Generation},
  author={Ti Zhou and Yanjie Zhao and Xinyi Hou and Xiaoyu Sun and Kai Chen and Haoyu Wang},
  journal={Proceedings of the ACM on Software Engineering},
  year={2025},
  volume={2},
  pages={219 - 241},
}

@inproceedings{Nguyen2015ReverseEM,
  title={Reverse Engineering Mobile Application User Interfaces with {REMAUI} ({T})},
  author={Tuan Anh Nguyen and Christoph Csallner},
  booktitle={Proceedings of the 30th {IEEE/ACM} International Conference on Automated Software Engineering ({ASE})},
  year={2015},
  pages={248--259},
  numpages={12},
  publisher={IEEE},
  address={Lincoln, NE, USA}
}

@article{Soselia2023LearningUR,
  title={Learning UI-to-Code Reverse Generator Using Visual Critic Without Rendering},
  author={Soselia, Davit and Saifullah, Khalid and Zhou, Tianyi},
  journal={arXiv preprint arXiv:2305.14637},
  year={2023},
  volume={abs/2305.14637},
  pages={1--10},
  numpages={10}
}

@article{Xu2021image2emmetAC,
  title={image2emmet: Automatic code generation from web user interface image},
  author={Yong Xu and Lili Bo and Xiaobing Sun and Bin Li and Jing Jiang and Wei Zhou},
  journal={Journal of Software: Evolution and Process},
  year={2021},
  volume={33},
  numpages = {12},
  pages={241-253},
}

@article{Wan2024AutomaticallyGU,
  title={Automatically Generating UI Code from Screenshot: A Divide-and-Conquer-Based Approach},
  author={Yuxuan Wan and Chaozheng Wang and Yi Dong and Wenxuan Wang and Shuqing Li and Yintong Huo and Michael R. Lyu},
  journal={ArXiv},
  year={2024},
  volume={abs/2406.16386},
 numpages = {12},
  pages={241-253}
}

@article{Wu2025MLLMBasedUA,
  title={MLLM-Based UI2Code Automation Guided by UI Layout Information},
  author={Fan Wu and Cuiyun Gao and Shuqing Li and Xinjie Wen and Qing Liao},
  journal={Proceedings of the ACM on Software Engineering},
  year={2025},
  volume={2},
  pages={1123 - 1145},
}

@article{Jiang2025ScreenCoderAV,
  title={ScreenCoder: Advancing Visual-to-Code Generation for Front-End Automation via Modular Multimodal Agents},
  author={Yilei Jiang and Yaozhi Zheng and Yuxuan Wan and Jiaming Han and Qunzhong Wang and Michael R. Lyu and Xiangyu Yue},
  journal={ArXiv},
  year={2025},
  volume={abs/2507.22827},
  pages = {1-20},
  numpages = {20},
}

@article{Laurenon2024UnlockingTC,
  title={Unlocking the conversion of Web Screenshots into HTML Code with the WebSight Dataset},
  author={Hugo Laurençon and L{\'e}o Tronchon and Victor Sanh},
  journal={ArXiv},
  year={2024},
  volume={abs/2403.09029},
  pages = {1-9},
  numpages = {9},
}

@article{Yang2025UIUGAU,
  title={UI-UG: A Unified MLLM for UI Understanding and Generation},
  author={Hao Yang and Weijie Qiu and Ru Zhang and Zhou Fang and Ruichao Mao and Xiaoyu Lin and Maji Huang and Zhaosong Huang and Teng Guo and Shuoyang Liu and Hai Rao},
  journal={ArXiv},
  year={2025},
  volume={abs/2509.24361},
  pages = {1-16},
  numpages = {16},
}

@article{Le2020DeepLF,
  title={Deep Learning for Source Code Modeling and Generation},
  author={Triet Huynh Minh Le and Hao Chen and Muhammad Ali Babar},
  journal={ACM Computing Surveys (CSUR)},
  year={2020},
  volume={53},
  pages={1 - 38},
}

@article{Li2023StarCoderMT,
  title={StarCoder: may the source be with you!},
  author={Raymond Li and Loubna Ben Allal and Yangtian Zi and others},
  journal={Trans. Mach. Learn. Res.},
  year={2023},
  volume={2023},
  pages = {1-55},
  numpages = {55},
}

@article{wei2024aiinspired,
  author={Wei, Jialiang and Courbis, Anne-Lise and Lambolais, Thomas and Dray, Gérard and Maalej, Walid},
  journal={IEEE Software}, 
  title={On AI-Inspired User Interface Design}, 
  year={2025},
  volume={42},
  number={3},
  pages={50-58},
  doi={10.1109/MS.2025.3536838}
}

@inproceedings{Lu2024MistyUP,
author = {Lu, Yuwen and Leung, Alan and Swearngin, Amanda and Nichols, Jeffrey and Barik, Titus},
title = {Misty: UI Prototyping Through Interactive Conceptual Blending},
year = {2025},
isbn = {9798400713941},
publisher = {Association for Computing Machinery},
address = {New York, NY, USA},
url = {https://doi.org/10.1145/3706598.3713924},
doi = {10.1145/3706598.3713924},
booktitle = {Proceedings of the 2025 CHI Conference on Human Factors in Computing Systems},
articleno = {1108},
numpages = {17},
location = {},
series = {CHI '25}
}

@inproceedings{Bunian2021VINSVS,
author = {Bunian, Sara and Li, Kai and Jemmali, Chaima and Harteveld, Casper and Fu, Yun and Seif El-Nasr, Magy Seif},
title = {VINS: Visual Search for Mobile User Interface Design},
year = {2021},
isbn = {9781450380966},
publisher = {Association for Computing Machinery},
address = {New York, NY, USA},
url = {https://doi.org/10.1145/3411764.3445762},
doi = {10.1145/3411764.3445762},
booktitle = {Proceedings of the 2021 CHI Conference on Human Factors in Computing Systems},
articleno = {423},
numpages = {14},
location = {Yokohama, Japan},
series = {CHI '21}
}

@article{her2025zeroshot3dmap,
  title={Zero-shot 3D Map Generation with LLM Agents: A Dual-Agent Architecture for Procedural Content Generation},
  author={Her, Lim Chien and Yan, Ming and Bai, Yunshu and Li, Ruihao and Zhang, Hao},
  journal={arXiv preprint arXiv:2512.10501},
  year={2025},
  volume={abs/2512.10501},
  pages={1--12},      
  numpages={12}
}
